\documentclass[sigconf, nonacm, authorversion]{acmart}

\pdfoutput=1

\usepackage{subfigure}
\usepackage{multirow}
\usepackage{tabularx}
\usepackage{kotex}
\usepackage[utf8]{inputenc}
\usepackage{caption} 
\usepackage[many]{tcolorbox}
\usepackage{array}
\usepackage{makecell}
\usepackage{soul}
\usepackage{setspace}
\newcolumntype{P}[1]{>{\centering\arraybackslash}p{#1}}
\newcolumntype{M}[1]{>{\centering\arraybackslash}m{#1}}

\newtcolorbox{FindingBox}{
    colback = white,
    boxsep = -6pt,
    top = 10pt,
    bottom = 10pt,
    arc = 0pt,
}

\AtBeginDocument{%
  }

\begin{document}

\title{Impact of Large Language Models of Code on Fault Localization}

\author{Suhwan Ji}
\orcid{0000-0001-8934-6099}
\affiliation{%
  \institution{Yonsei University}
  \city{Seoul}
  \country{Republic of Korea}
}
\email{shji@yonsei.ac.kr}

\author{Sanghwa Lee}
\orcid{0000-0002-0973-7784}
\affiliation{%
  \institution{Kangwon National University}
  \city{Chuncheon}
  \country{Republic of Korea}
}
\email{lion0738@kangwon.ac.kr}

\author{Changsup Lee}
\orcid{0009-0007-8887-7614}
\affiliation{%
  \institution{Kangwon National University}
  \city{Chuncheon}
  \country{Republic of Korea}
}
\email{cslee@kangwon.ac.kr}

\author{Hyeonseung Im}
\orcid{0000-0002-3901-0834}
\authornote{Corresponding author.}
\affiliation{%
  \institution{Kangwon National University}
  \city{Chuncheon}
  \country{Republic of Korea}
  }
\email{hsim@kangwon.ac.kr}

\author{Yo-Sub Han}
\orcid{0000-0002-7211-6657}
\affiliation{%
  \institution{Yonsei University}
  \city{Seoul}
  \country{Republic of Korea}
  }
\email{emmous@yonsei.ac.kr}

\renewcommand{\shortauthors}{Ji et al.}

\begin{abstract}
Identifying the point of error is imperative in software debugging. Traditional fault localization (FL) techniques rely on executing the program and using the code coverage matrix in tandem with test case results to calculate a suspiciousness score for each function or line. Recently, learning-based FL techniques have harnessed machine learning models to extract meaningful features from the code coverage matrix and improve FL performance. These techniques, however, require compilable source code, existing test cases, and specialized tools for generating the code coverage matrix for each programming language of interest. 

In this paper, we propose, for the first time, a simple but effective sequence generation approach for fine-tuning large language models of code (LLMCs) for FL tasks. LLMCs have recently received much attention for various software engineering problems such as code completion, summarization, translation, and refinement. In line with these, we leverage the innate understanding of code that LLMCs have acquired through pre-training on large code corpora. Specifically, we fine-tune representative encoder, encoder-decoder, and decoder-based 13 LLMCs (across 7 different architectures) for FL tasks. Unlike previous approaches, LLMCs can analyze code sequences even with syntactic errors, since they do not rely on compiled input. Still, they have a limitation on the length of the input data. Therefore, for a fair comparison with existing FL techniques, we extract methods with errors from the project-level benchmark, Defects4J, and analyze them at the line level. Experimental results show that LLMCs fine-tuned with our approach successfully pinpoint error positions in 50.6\%, 64.2\%, and 72.3\% of 1,291 methods in Defects4J for Top-1/3/5 prediction, outperforming the best learning-based state-of-the-art technique by up to 1.35, 1.12, and 1.08 times, respectively. We also conduct an in-depth investigation of key factors that may affect the FL performance of LLMCs. Our findings suggest promising research directions for FL and automated program repair tasks using LLMCs.
\end{abstract}



\maketitle

\section{Introduction}
\label{sec:intro}

Programs are likely to contain unintentional defects due to a lack of understanding of the specifications or carelessness, making program debugging essential. To debug a program, one must first locate where the defect occurs and then modify that part to ensure it no longer manifests. Developers spend roughly half their development time debugging to fix unintended defects~\cite{grams2019much}. To assist developers in the debugging process, much research has been proposed on automatically fixing defects. For example, Jiang~\emph{et al.}~\cite{jiang2023impact} and Huang~\emph{et al.}~\cite{huang2023empirical} conducted research on fine-tuning large language models of code (LLMCs) pre-trained on massive code datasets to automatically fix program defects. However, their focus is on fixing defects assuming the fault locations are already known, and thus this approach is difficult to use when the fault location is unknown.

Traditional fault localization (FL) techniques often construct a coverage matrix by executing the program with its input/output test data~\cite{wong2016survey}. This matrix records the code lines traversed by failing and passing test cases. The defect suspiciousness of each line is then calculated from the matrix using similarity formulas (\emph{e.g.}, cosine similarity). Such traditional FL techniques can be categorized into spectrum-based fault localization (SBFL)~\cite{abreu2009practical, jones2002visualization, abreu2006evaluation, wong2007effective, abreu2007accuracy, abreu2009spectrum, naish2011model, yoo2012evolving, wong2013dstar} and mutation-based fault localization (MBFL)~\cite{moon2014ask, papadakis2015metallaxis, tufano2020deepmutation, tian2022learning}. SBFL executes the original program code, while MBFL mutates small code snippets like operators to populate the coverage matrix with more information, aiming to improve FL performance. However, SBFL and MBFL require many test cases to build the coverage matrix, and MBFL is particularly time-consuming due to the mutation operations.

As deep learning research has gained traction, learning-based FL (LBFL) techniques utilizing deep learning models have also emerged~\cite{zhang2021improving, b2016learning, zhang2017deep}. LBFL employs complex models like convolutional neural networks (CNNs) or graph-based CNNs (GCNs) to extract intricate information from the coverage matrix that similarity formulas can not capture, enabling the localization of program defects~\cite{zhang2019cnn,lou2021boosting}. Some LBFL approaches also extract textual similarity information from the coverage matrix or train models with additional data such as defect types~\cite{li2019deepfl,meng2022improving}. However, LBFL models are limited in that obtaining not only the coverage matrix but also the additional information needed for training is time-consuming. Moreover, there are cases where the results of these models are not reproducible~\cite{meng2022improving,lou2021boosting}, raising concerns about their reliability and practical applicability.

Recently, with rapidly advancing computing resources, large language models (LLMs) based on the Transformers~\cite{vaswani2017attention} with attention mechanisms have been utilized in various domains. In particular, for program-related tasks, many pre-trained models have been proposed~\cite{feng2020codebert,guo2020graphcodebert,wang2021codet5,ahmad2021unified,guo2022unixcoder,nijkamp2022codegen,fried2022incoder} that use massive unlabeled natural language and source code data~\cite{husain2019codesearchnet,kocetkov2022stack} from sources like GitHub and Stack Overflow for pre-training. These pre-trained models have demonstrated impressive performance on various downstream tasks, such as code generation~\cite{zheng2023codegeex}, code summarization~\cite{guo2022unixcoder}, and code search~\cite{wang2023codet5+}, after fine-tuning. LLMCs have the advantage of enabling models to understand programs from structural information by utilizing pre-training methods~\cite{guo2020graphcodebert, guo2022unixcoder, niu2022spt} that, unlike natural language, incorporate programs' abstract syntax trees (ASTs) or data flow graphs (DFGs). This structural awareness allows LLMCs to be effectively applied across diverse downstream tasks. 

To address the time and reproducibility limitations of existing FL studies, Yang~\emph{et al.}~\cite{yang2024large} propose LLMAO, a test-free FL technique based on CodeGen~\cite{nijkamp2022codegen}, which is an LLMC that learns strong program understanding and representation capabilities from source code sequences. Since fine-tuning LLMCs requires substantial computing resources, the authors employ parameter-efficient fine-tuning (PEFT)~\cite{ding2023parameter} to effectively leverage these models. Specifically, instead of directly fine-tuning pre-trained LLMCs, they add a compact bidirectional adapter model on top of a frozen CodeGen model. They embed the input source code into well-represented vectors using CodeGen, from which special tokens representing each line of code are extracted, and train the bidirectional adapter using these token sequence data. This approach enables memory-efficient FL while leveraging the pre-trained knowledge of LLMCs. However, since LLMAO does not fine-tune the LLMCs themselves, there is potential for further performance gains through direct model adaptation to the FL task.

In this paper, we propose a simple yet effective sequence generation (SG) approach to directly fine-tuning pre-trained LLMCs for FL. To our knowledge, this is the first work to fine-tune LLMCs for this task. We fine-tune 13 LLMCs across 7 different architectures and compare their FL performance across various defect types. Unlike existing FL techniques that require executing the target programs, our approach, similar to LLMAO, allows for the complete separation of training and evaluation data, as there is no need for program execution. Our fine-tuned LLMCs demonstrate superior performance compared to the state-of-the-art (SOTA) FL technique~\cite{meng2022improving}, which, unlike ours, is evaluated using $N$-fold cross-validation (CV) without data separation. Moreover, we find that LLMCs with encoder architectures that effectively represent input sequences by leveraging structural information, or those pre-trained considering bidirectional context, perform better on FL. Unlike simple SG, FL involves structural program data, making such architectural choices and bidirectional pre-training beneficial.

The main contributions are summarized as follows.
\begin{itemize}
\item 
To the best of our knowledge, we propose, for the first time, a simple but effective SG approach to fine-tuning different types of LLMCs for FL. With our method, the performance of encoder, encoder-decoder, and decoder-based LLMCs is improved by up to 1.72, 1.65, and 1.77 times, respectively, in terms of the Top-1 metric.

\item 
We conduct a comprehensive evaluation of seven LLMCs (i.e. CodeBERT, GraphCodeBERT, PLBART, CodeT5, UniXcoder, CodeGen, and InCoder) fine-tuned using our approach, in comparison with four traditional and three SOTA learning-based methods. Experimental results show that UniXcoder, which shows the best performance, is 1.35/1.12/1.08 times better than the best SOTA method on the Defects4J benchmark with respect to the Top-1/3/5 metric.

\item 
We also conduct an in-depth investigation of key factors that may affect the performance of LLMCs, such as their architecture and size, problem domain (i.e. logical error and vulnerability detection), programming languages of interest, and the use of pre-training and line-numbering. 

\item
To ensure reproducibility of the results, the source code and data used in this paper will be made publicly available soon.
\end{itemize}

\section{Preliminaries}
\label{sec:preliminaries}

\subsection{Large Language Models}

The core of LLMs is the Transformer~\cite{vaswani2017attention}, an attention-based model with an encoder-decoder architecture. It allows fast training through parallel computation over tokens in a sequence. The Transformer has shown excellent performance in neural machine translation tasks. Building upon this model, several LLMs have emerged~\cite{lewis2019bart, raffel2020exploring}, pre-trained on massive datasets. In addition to the encoder-decoder structure, derived architectures like encoder-only and decoder-only architectures have also been developed~\cite{devlin2018bert, achiam2023gpt}.

LLMs have demonstrated impressive performance across most natural language processing tasks~\cite{zhao2023survey}, inspiring their applications in various domains, including software engineering (SE)~\cite{fan2023large, jiang2024survey}. This is because software shares many statistical properties with natural language~\cite{devanbu2015new}. Consequently, LLMs trained on code are expected to perform well on SE tasks, learning not only natural language and code but also ASTs and DFGs.

\subsection{Encoder-based Pre-trained Models}

Encoder-based models allow bidirectional attention, enabling them to use context from both directions and compress information from whole sentences. However, encoders alone cannot generate outputs; they extract essential features and pass them to generative models. Encoder-based pre-trained models like BERT~\cite{devlin2018bert} have been developed using methods such as masked language modeling and next sentence prediction. CodeBERT~\cite{feng2020codebert} is the first to be pre-trained on both natural language and programming languages, which has proven effective for code search and code-to-text generation. However, treating code as a sequence of tokens poses challenges in capturing its inherent structural properties. GraphCodeBERT~\cite{guo2020graphcodebert} addresses this by jointly pre-training on DFGs, demonstrating its effectiveness on code-related tasks. Other encoder-based pre-trained models include CuBERT~\cite{kanade2020learning} and PolyglotCodeBERT~\cite{ahmed2022multilingual}.

\subsection{Decoder-based Pre-trained Models}

Decoder-based models use only unidirectional attention, generating the next token based on previous tokens to construct sentences. Decoder-based pre-trained models like GPT~\cite{radford2018improving} have emerged, using causal language modeling for pre-training. Generating code matching the user's intent in a single turn is challenging. CodeGen~\cite{nijkamp2022codegen} tackles this problem by multi-turn generation with a larger model. InCoder~\cite{fried2022incoder} proposes a decoder-based approach using bidirectional context and proves its effectiveness for code completion and refinement. Other decoder-based pre-trained models include Codex~\cite{chen2021evaluating}, CodeGeeX~\cite{zheng2023codegeex}, and StarCoder~\cite{li2023starcoder}.

\subsection{Encoder-Decoder-based Pre-trained Models}

Encoder-decoder-based models combine the bidirectional understanding capabilities of encoders with the unidirectional generation capabilities of decoders. They either combine separately trained encoders and decoders or jointly train the entire structure, using pre-training methods such as denoising auto-encoding and text-to-text tasks. PLBART~\cite{ahmad2021unified} uses a denoising auto-encoding to reconstruct original data from noisy inputs and demonstrates strong performance in code summarization, generation, and translation. CodeT5~\cite{wang2021codet5} takes a text-to-text approach, where the encoder receives a request and the decoder generates a response, allowing a unified approach across tasks and achieving impressive results on various SE tasks. UniXcoder~\cite{guo2022unixcoder} shares the same weights between the encoder and decoder, pre-training jointly on ASTs and comments in a multi-modal fashion. Other encoder-decoder-based pre-trained models include AlphaCode~\cite{li2022competition} and CodeT5+~\cite{wang2023codet5+}.

\section{Our Approach}
\label{sec:approach}

\begin{figure*}[t]
    \centering
    \includegraphics[width=\textwidth]{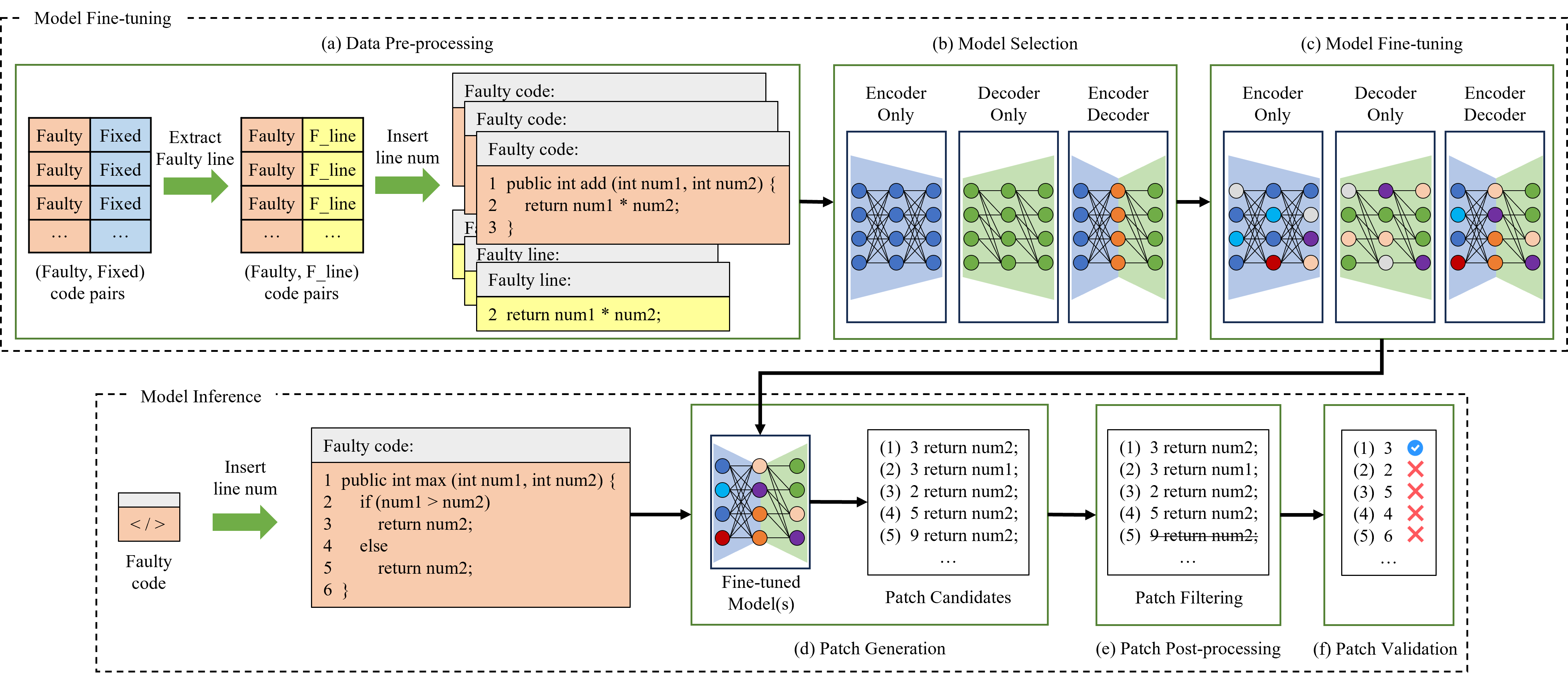}
    \caption{Workflow for FL based on fine-tuning LLMCs using our approach}
    \label{fig:workflow}
\end{figure*}

Transformer-based LLMCs, pre-trained with large-scale code data, learn the ability to understand code during the pre-training process. When the pre-training method and a downstream task of interest differ, the LLMCs need to be fine-tuned for the specific task to take advantage of the pre-trained capability. In this paper, we propose to add line numbers to the input source code before performing fine-tuning to predict faulty lines in the code. The line numbers serve as an important hint for LLMCs to leverage their pre-trained capability for the FL task. Figure~\ref{fig:workflow} illustrates the overall workflow of fine-tuning LLMCs for FL. Specifically, we first extract short functions (methods) from a long program at the project level and add line numbers to the functions. Second, we fine-tune an LLMC to learn to perform FL by generating sequences using these pre-processed data. Third, we create a patch of faulty lines from benchmark datasets using the fine-tuned model. Lastly, the performance of the model is evaluated by comparing the generated patches with the actual faulty locations. Each step is explained in more detail below.

\subsection{Data Pre-processing}

LLMCs are limited in the number of tokens that can be processed at one time, depending on the model design and memory efficiency~\cite{guo2023longcoder}. Therefore, it is impossible to analyze a long project-level program at once using LLMCs. To address this problem, a long program can be summarized and converted into a short program, or a sliding window technique can be used to divide the program into smaller sections and then combine the individual results~\cite{zaheer2020big, yang2024large}. Instead, in this paper, we focus only on analyzing relatively short functions with errors extracted from larger, defective programs at the project level. LLMCs can then estimate a faulty line from the input sequence simply by generating the faulty line as it is. However, this method has the disadvantage of making it difficult to correctly estimate the faulty line if the same code snippet appears elsewhere in the function. To solve this problem, we propose a simple but effective method of adding a line number (and a tab symbol) at the beginning of each line. This allows LLMCs to uniquely identify each line and locate the faulty line.

\subsection{Fine-tuning}

\begin{table*}[t]
  \caption{Details of the LLMCs considered in this study}
  \label{tbl:llmcs-details}
  \small
  \begin{tabular}{ccccc} 
    \toprule
    \textbf{Model (Year)} & \textbf{Size} & \textbf{Context Length} & \textbf{Type} & \textbf{Pre-training Dataset} \\
    \midrule
    CodeBERT (2020)~\cite{feng2020codebert}             & 125M & 512 & Encoder only & CodeSearchNet~\cite{husain2019codesearchnet} \\ 
    GraphCodeBERT (2020)~\cite{guo2020graphcodebert}    & 125M & 512 & Encoder only & CodeSearchNet \\ 
    PLBART (2021)~\cite{ahmad2021unified}               & 140M/400M & 512 & Encoder-Decoder & Stack Overflow, BigQuery \\ 
    CodeT5 (2021)~\cite{wang2021codet5}                 & 60M/215M/710M & 512 & Encoder-Decoder & CodeSearchNet, BigQuery \\ 
    UniXcoder (2022)~\cite{guo2022unixcoder}           & 125M & 768 & Encoder-Decoder & C4~\cite{raffel2020exploring}, CodeSearchNet \\ 
    CodeGen (2023)~\cite{nijkamp2022codegen}           & 350M/2.6B/6.6B & 768 & Decoder only & ThePile~\cite{gao2020pile}, BigQuery, BigPython \\ 
    InCoder (2023)~\cite{fried2022incoder}              & 1.2B/6.2B & 1024 & Decoder only & Github, Gitlab, Stack Overflow \\
    \bottomrule
  \end{tabular}
\end{table*}
   
To identify LLMCs suitable for FL, we fine-tune 13 pre-trained models (across 7 different architectures), as shown in Table~\ref{tbl:llmcs-details}, using the sequence generation (SG) method. Since the SG method uses token sequences of code as input and output, LLMCs need to have a decoder structure to generate a sequence. Therefore, as for CodeBERT~\cite{feng2020codebert} and GraphCodeBERT~\cite{guo2020graphcodebert}, which are of encoder-only structure, we create a sequence-to-sequence structure by connecting six decoder layers of a randomly initialized Transformer~\cite{vaswani2017attention} behind the model to perform FL by generating a sequence. Since encoder-decoder and decoder-only models use sequences as input and output, the rest of the models in Table~\ref{tbl:llmcs-details} are fine-tuned without modification.

More specifically, LLMCs are fine-tuned using SG for the FL task as follows. Suppose a faulty pre-processed code $S$ has $n$ lines and a fault in the $k$-th line. Then, a pre-trained LLMC $\mathcal{M}$ is fine-tuned to predict the fault location $k$ and the code fragment at the $k$-th line as it is. That is, $\mathcal{M}(S) = O$, where $S = \{1\ l_1,\ 2\ l_2,\ ...,\ k\ l_k,\ ...,\ n\ l_n\}$. Here, $m$ and $l_m$ denote a line number and the code fragment at the $m$-th line, respectively (with $m\ l_m$ representing a token sequence converted to $t$ tokens by the LLMC tokenizer). Additionally, $O = k\ l_k$, where $k$ and $l_k$ denote the fault location and the corresponding code fragment.

\subsection{Patch Generation and Validation}

Fine-tuned LLMCs take a token sequence of faulty code and return the fault location and a token sequence of the corresponding line. A model with Transformer's decoder structure generates a sequence of tokens in an auto-regressive manner, that is, it generates a token from the input sequence and adds it to the end of the input sequence to predict the next token~\cite{vaswani2017attention}. In this approach, since the tokens generated initially have a significant impact on the overall generation result, beam search~\cite{graves2012sequence} is used to generate token sequences with diverse patterns, which in our context correspond to various candidate patches at multiple locations.

To evaluate FL performance, we use the Top-$N$ metric~\cite{re2006efficient}. If the fault location in the code is included among the Top-$N$ patch candidates generated by an LLMC, the model is considered to have correctly identified the fault location within the Top-$N$. Patch candidates are sorted by their probability of occurrence, and verifying the patches is straightforward due to the line-numbering technique. Since the patch is of the form ($k\backslash t\ l_k$), $k$ before $\backslash t$ is compared to the actual fault location.

\section{Experimental Design}
\label{sec:experimental-design}

\subsection{Research Questions}

We explore the FL capability of LLMCs in various scenarios by answering the following research questions.
\begin{itemize}
\item 
\textbf{RQ1. How do different design choices affect the FL capability of LLMCs?}
We fine-tune representative encoder-, encoder-decoder-, and decoder-based LLMCs to determine which architecture is best suited for FL. We also investigate the impact of LLMCs' parameter size on FL capability.
\item
\textbf{RQ2. How effective are LLMCs compared to the state-of-the-art FL techniques?}
We compare the performance of LLMCs with existing traditional and SOTA learning-based FL techniques to see if LLMCs are superior.
\item
\textbf{RQ3. How well do LLMCs perform on various FL benchmarks?}
We evaluate the performance of LLMCs on various FL benchmarks to assess their applicability to different languages and problem domains (i.e. logical error and vulnerability detection).
\item
\textbf{RQ4. How does pre-training or line-numbering technique affect the FL performance of LLMCs?}
We conduct ablation studies that exclude either of the two components applied to the LLMCs.
\end{itemize}

\subsection{Dataset}
Traditional fault localization tools typically measure code coverage information from which fault suspiciousness scores are calculated~\cite{campos2012gzoltar}. Given a program, to obtain enough code coverage information to calculate its fault suspiciousness, it is necessary to have test cases that trigger the fault as well as successful test cases that the program passes. Having a larger number of test cases typically leads to increased code coverage, which can improve the accuracy of fault suspiciousness calculations~\cite{baudry2006improving}. However, it is difficult to collect faulty programs, and it requires significant effort to construct passing and failing test cases. Because of these limitations, previous research~\cite{li2019deepfl, lou2021boosting, wu2023gmbfl, yang2024large} has often evaluated FL performance using $N$-fold CV on the widely-used Defects4J~\cite{just2014defects4j} benchmark. 

Unlike the traditional approach, which relies on code coverage information obtained by executing test cases, this study utilizes LLMCs for FL by leveraging the knowledge acquired through pre-training. To evaluate the effectiveness of LLMCs for FL, we fine-tune them on Java, Python, and C/C++ benchmarks, as summarized in Table~\ref{tbl:datasets}, by applying the line-numbering technique to the input source code without utilizing any other code metrics.

\begin{table}[t]
  \caption{Details of fine-tuning and evaluation datasets}
  \label{tbl:datasets}
  \small
  \setlength{\tabcolsep}{2pt}
  \begin{tabular}{l|c|cc|c|c|c|c}
    \toprule
    \multirow{3}{*}{Type} & \multicolumn{5}{c|}{Java} & Python & C/C++ \\ \cmidrule{2-8}
    & \multicolumn{1}{c|}{\multirow{2}{*}{Recoder}} & \multicolumn{2}{c|}{Defects4J} & \multicolumn{1}{c|}{\multirow{2}{*}{HumanEval}} & \multicolumn{1}{c|}{\multirow{2}{*}{QuixBugs}} & \multirow{2}{*}{CodeNet} & \multirow{2}{*}{\makecell{CVEfixes\\ 5-fold}} \\
    & & \multicolumn{1}{c}{v1.2} & \multicolumn{1}{c|}{v2.0} & & & & \\ 
    \midrule
    Train & 129,310 & - & - & - & - & 96,000 & 952 \\
    Valid & 14,356 & - & - & - & - & 12,000 & - \\
    Test & - & 535 & 756 & 164 & 40 & 12,000 & 238 \\ 
    \bottomrule   
    \end{tabular}
\end{table}

\textbf{Java fine-tuning.}
To optimize LLMCs as an FL model for Java, we fine-tune them using the training dataset of Recoder~\cite{zhu2021syntax}, an automated program repair (APR) model. The dataset consists of (faulty code, correct code) pairs constructed from GitHub Java projects created between 2011 and 2018. To avoid data leakage, it excludes not only projects cloning Defects4J v1.2 and v2.0, but also projects fixing faults in the same way as Defects4J~\cite{just2014defects4j}, identified through abstract syntax tree comparison. In this study, we use 129,310 pairs from Recoder for fine-tuning and 14,356 pairs as validation data.

\textbf{Java evaluation.}
To evaluate the FL performance of LLMCs fine-tuned using our approach and to compare it with previous studies, we mainly use the Defects4J benchmark~\cite{just2014defects4j} containing real defects from popular open-source Java projects. It includes faulty programs, their corresponding fixes, and test cases triggering the defects. Most studies utilize the 395 defects from version 1.2 and the additional 444 defects from version 2.0. In this study, we also use the same versions for comparison with previous approaches. However, as Defects4J consists of long project-level programs and LLMCs have input sequence length limitations, we extract faulty methods for evaluation (v1.2 = 535, v2.0 = 756).

In addition, we use the HumanEval-Java and QuixBugs benchmarks. HumanEval-Java~\cite{jiang2023impact} is a conversion of the HumanEval~\cite{chen2021evaluating} benchmark from Python to Java, designed for evaluating ARP performance. It consists of manually created, complex bugs (single hunks) spanning multiple lines. HumanEval is a human-curated Python benchmark with unique data not included in pre-training, used to assess the code generation performance of LLMCs. This benchmark serves as an important tool for evaluating and comparing the code generation and error correction capabilities of different LLMCs (N = 164).

QuixBugs~\cite{lin2017quixbugs} is a benchmark comprising 40 algorithmic problems written in Java and Python from the Quixey Challenge, each containing a single line bug. It is primarily used to evaluate the effectiveness of APR tools across multiple languages~\cite{zhang2023survey} (N = 40).

\textbf{Python.}
CodeNet~\cite{puri2021codenet} is a large-scale benchmark designed to drive AI innovation in code-related tasks. It comprises approximately 14 million code samples for over 4,000 algorithmic problems, collected from the online judge platforms AIZU and AtCoder. In this study, we use Python data where the difference between correct and incorrect solutions is limited to a single line (120,000 data are split in an 8:1:1 ratio for train/validation/test sets).

\textbf{C/C++.}
To evaluate vulnerability detection performance, we utilize the CVEfixes~\cite{bhandari2021cvefixes} dataset, which contains C/C++ common vulnerabilities and exposures (CVEs) from the U.S. National Vulnerability Database. Due to the limited availability of curated vulnerability data for fine-tuning, we perform 5-fold CV on 1,190 data with a maximum of 2 line differences between vulnerable and patched code snippets.

\subsection{Compared Techniques}

We compare our approach with seven existing FL techniques: two spectrum-based (Ochiai~\cite{abreu2006evaluation}, Jaccard~\cite{abreu2009practical}), two mutation-based ( MUSE~\cite{moon2014ask}, Metallaxis~\cite{papadakis2015metallaxis}), three learning-based (DeepFL~\cite{li2019deepfl}, XAI4FL~\cite{widyasari2022xai4fl}, TRANSFER-FL~\cite{meng2022improving}). Additionally, we evaluate our approach against a recent parameter-efficient fine-tuning (PEFT)-based technique called LLMAO~\cite{yang2024large}, specifically tailored for CodeGen~\cite{nijkamp2022codegen}.

Table~\ref{tbl:fl-methods} describes how we compare our approach to existing FL techniques. Spectrum-based features for SBFL are generated using GZoltar~\cite{campos2012gzoltar}, while mutation-based features for MBFL are generated using Major~\cite{just2014major}. For DeepFL and TRANSFER-FL, the same spectrum-based features are used, while the mutation-based features are generated using PITest~\cite{coles2016pit}, as in their reference implementations. However, some defects in Defects4J v2.0 are impossible to reproduce, as noted in~\cite{meng2022improving, lou2021boosting}, and these are classified as missed faults for performance comparison. Meanwhile, DeepFL detects faulty functions using various function-level information. However, since our focus is on detecting faults within functions at the line level, the DeepFL model is implemented without some function-level features such as textual similarity and code complexity, as in previous studies~\cite{meng2022improving, li2021fault}. XAI4FL and TRANSFER-FL are used without modification and tested on Defects4J including v2.0.

Existing FL techniques compute a suspiciousness score for each line of code in large project-level programs, but LLMCs have input sequence length limitations. Therefore, for LLMCs, we extract and analyze individual faulty functions from the project-level programs. For comparison, function-level suspiciousness scores are derived from the project-level scores computed by existing techniques. More specifically, among the project-level suspiciousness rankings, only the lines within the same faulty functions are considered for evaluating the Top-$N$.

To compare our approach with LLMAO, which was originally trained on project-level programs, we consider its three variants. First, `LLMAO (10-fold)' performs 10-fold CV on Defects4J v1.2 as done in~\cite{yang2024large} but at the function level. Second, `LLMAO (new projects)' evaluates the model trained on function-level Defects4J v1.2 on function-level Defects4J v2.0. Lastly, `LLMAO (Recoder)' fine-tunes LLMAO on the Recoder data and evaluates it on both Defects4J v1.2 and v2.0 at the function level. All LLMAO variants use the CodeGen-16B model as the underlying model.

\begin{table*}[]
\caption{Summary of compared FL techniques}
\label{tbl:fl-methods}
\setlength{\tabcolsep}{3pt}
\small
\begin{tabular}{c|c|c|l|l}
\toprule
\textbf{Category} & \textbf{Employed tools} & \textbf{FL technique} & \multicolumn{1}{c|}{\textbf{Original method}} & \multicolumn{1}{c}{\textbf{Modification for comparison}} \\ \midrule
SG-based & \multirow{2}{*}{LLMCs} & Ours & Suspect each line within the faulty function & - \\ \cline{1-1} \cline{3-5} 
PEFT-based &  & LLMAO~\cite{yang2024large} & \multirow{7}{*}{Suspect each line within the faulty project} & \multirow{7}{*}{Suspect each line within the faulty function} \\ \cline{1-3}
\multirow{2}{*}{Spectrum-based} & \multirow{2}{*}{GZoltar~\cite{campos2012gzoltar}} & Ochiai~\cite{abreu2006evaluation} &  &  \\ \cline{3-3}
 &  & Jaccard~\cite{abreu2009practical} &  &  \\ \cline{1-3}
\multirow{2}{*}{Mutation-based} & \multirow{2}{*}{Major~\cite{just2014major}} & MUSE~\cite{moon2014ask} &  &  \\ \cline{3-3}
 &  & Metallaxis~\cite{papadakis2015metallaxis} &  &  \\ \cline{1-3}
\multirow{5}{*}{Learning-based} & GZoltar, PITest~\cite{coles2016pit} & TRANSFER-FL~\cite{meng2022improving} &  &  \\ \cline{2-3}
 & GZoltar & {XAI4FL}~\cite{widyasari2022xai4fl} &  &  \\ \cline{2-5} 
 & GZoltar, PITest & DeepFL~\cite{li2019deepfl} & \makecell[l]{Suspect each function within the faulty \\ project and use spectrum, mutation and \\ function features} & \makecell[l]{Suspect each line within the faulty function \\ and use only spectrum and mutation features} \\ 
\bottomrule
\end{tabular}
\end{table*}

\subsection{Environment}

As shown in Table~\ref{tbl:llmcs-details}, we used 7 different LLMCs with a total of 13 instances, varying in parameter sizes. For fine-tuning, we used four NVIDIA GeForce RTX 4090 GPUs (one for $\le$ 750M, two for $\le$ 2B, and four for 6B parameter models). The batch size was set to the maximum possible without causing out-of-memory errors. With abundant training data and rapid convergence, we fine-tuned LLMCs for only 1 epoch (except for C/C++ data, where 5 epochs were used to mitigate under/over-fitting due to the smaller dataset size). For accurate performance comparison, we created five individual fine-tuned models for each LLMC instance and measured the average performance of the top three models.

\section{Empirical Evaluation}
\label{sec:evaluation}

\subsection{Impacts of Design Choices of LLMCs}

\begin{table*}[t]
  \caption{RQ1. FL performance of LLMCs of various types and sizes fine-tuned by our approach using Recoder and tested on the Defects4J benchmark $(N = 1,291)$}
  \label{tbl:RQ1-llmcs-types-sizes}
  \small
  \begin{tabular}{c|c|c|c|c|c|c|c|c|c|c|c|c|c} 
    \toprule
    \multirow{3}{*}{Metric} & \multicolumn{2}{c|}{\textbf{Encoder only}} & \multicolumn{6}{c|}{\textbf{Encoder-Decoder}} & \multicolumn{5}{c}{\textbf{Decoder only}} \\ \cmidrule{2-14}
    & \textbf{CodeBERT} & \textbf{GraphCodeBERT} & \multicolumn{2}{c|}{\textbf{PLBART}} & \multicolumn{3}{c|}{\textbf{CodeT5}} & \textbf{UniXcoder} & \multicolumn{3}{c|}{\textbf{CodeGen}} & \multicolumn{2}{c}{\textbf{InCoder}} \\
    & base & base & \multicolumn{1}{c}{base} & \multicolumn{1}{c|}{large} & \multicolumn{1}{c}{small} & \multicolumn{1}{c}{base} & \multicolumn{1}{c|}{large} & base & \multicolumn{1}{c}{350M} & \multicolumn{1}{c}{2B} & \multicolumn{1}{c|}{6B} & \multicolumn{1}{c}{1B} & \multicolumn{1}{c}{6B} \\
    \midrule
    Top-1 & 436.3 & 548.3 & \multicolumn{1}{c}{533.3} & \multicolumn{1}{c|}{530.6} & \multicolumn{1}{c}{513.6} & \multicolumn{1}{c}{520.6} & \multicolumn{1}{c|}{530.3} & \textbf{653.3} & \multicolumn{1}{c}{567} & \multicolumn{1}{c}{549} & \multicolumn{1}{c|}{550} & \multicolumn{1}{c}{579.3} & \multicolumn{1}{c}{552} \\ 
    Top-3 & 581.3 & 729.3 & \multicolumn{1}{c}{687.3} & \multicolumn{1}{c|}{695} & \multicolumn{1}{c}{675.3} & \multicolumn{1}{c}{676} & \multicolumn{1}{c|}{674.6} & \textbf{828.6} & \multicolumn{1}{c}{729} & \multicolumn{1}{c}{714} & \multicolumn{1}{c|}{720} & \multicolumn{1}{c}{772.6} & \multicolumn{1}{c}{736.3} \\
    Top-5 & 676 & 838.3 & \multicolumn{1}{c}{775.6} & \multicolumn{1}{c|}{784} & \multicolumn{1}{c}{769.6} & \multicolumn{1}{c}{770.6} & \multicolumn{1}{c|}{754} & \textbf{933.3} & \multicolumn{1}{c}{823} & \multicolumn{1}{c}{818} & \multicolumn{1}{c|}{815} & \multicolumn{1}{c}{886.3} & \multicolumn{1}{c}{853.3} \\
    \bottomrule
  \end{tabular}
\end{table*}

\textbf{RQ1. How do different design choices affect the FL capability of LLMCs?} Table~\ref{tbl:RQ1-llmcs-types-sizes} shows the FL performance of 13 LLMCs, fine-tuned using the Recoder dataset and our line-numbering technique, when tested on 1,291 Defects4J methods with defects.

\textbf{Model type.}
First, when comparing the FL performance by LLMC type, UniXcoder of encoder-decoder type shows the best performance by finding 653.3 (50.6\%), 828.6 (64.2\%), and 933.3 (72.3\%) defects (Top-1, 3, 5). More specifically, it was pre-trained using a masking technique that incorporates the learning methods of an encoder, decoder, and encoder-decoder into a single model using AST data. By learning to understand code in these three complementary ways, UniXcoder demonstrates the best performance. In contrast, PLBART (large) and CodeT5 (large), pre-trained solely using the encoder-decoder method, exhibit limited performance. The decoder-only types, CodeGen (350M) and InCoder (1B), show the performance of 43.9$\sim$44.9\%, 56.5$\sim$60\%, and 63.7$\sim$68.7\% (Top-1, 3, and 5), respectively. Between the two models, InCoder performs slightly better because it was trained to mask the middle of the input sequence and infer the masked part from the end of the sequence, thus better utilizing bidirectional information of the code. The encoder-only types, CodeBERT and GraphCodeBERT, show performance in the middle range compared to the encoder-decoder and decoder-only types, except for UniXcoder. (We note here that a randomly initialized decoder layer is added to CodeBERT and GraphCodeBERT to enable sequence generation for the FL task.) In particular, GraphCodeBERT takes both the code sequence and the DFG as input, utilizing the structural features of the code. Thus, it performs about 1.25 times better than CodeBERT and shows similar performance to CodeGen. Considering when each model was proposed, it is important to use a pre-trained encoder that effectively represents code for code tasks.

\begin{FindingBox}
\textbf{Finding 1:} For FL, LLMCs pre-trained to utilize bidirectional code information perform well. For example, UniXcoder of encoder-decoder type takes both code sequences and ASTs as input and was pre-trained using three types of learning methods. The encoder-only GraphCodeBERT also utilizes code sequences and DFGs. Moreover, the decoder-only InCoder was pre-trained via an infilling approach that incorporates bidirectional context by predicting the middle part.
\end{FindingBox}

\begin{figure}[t]
    \centering
    \includegraphics[width=0.45\textwidth]{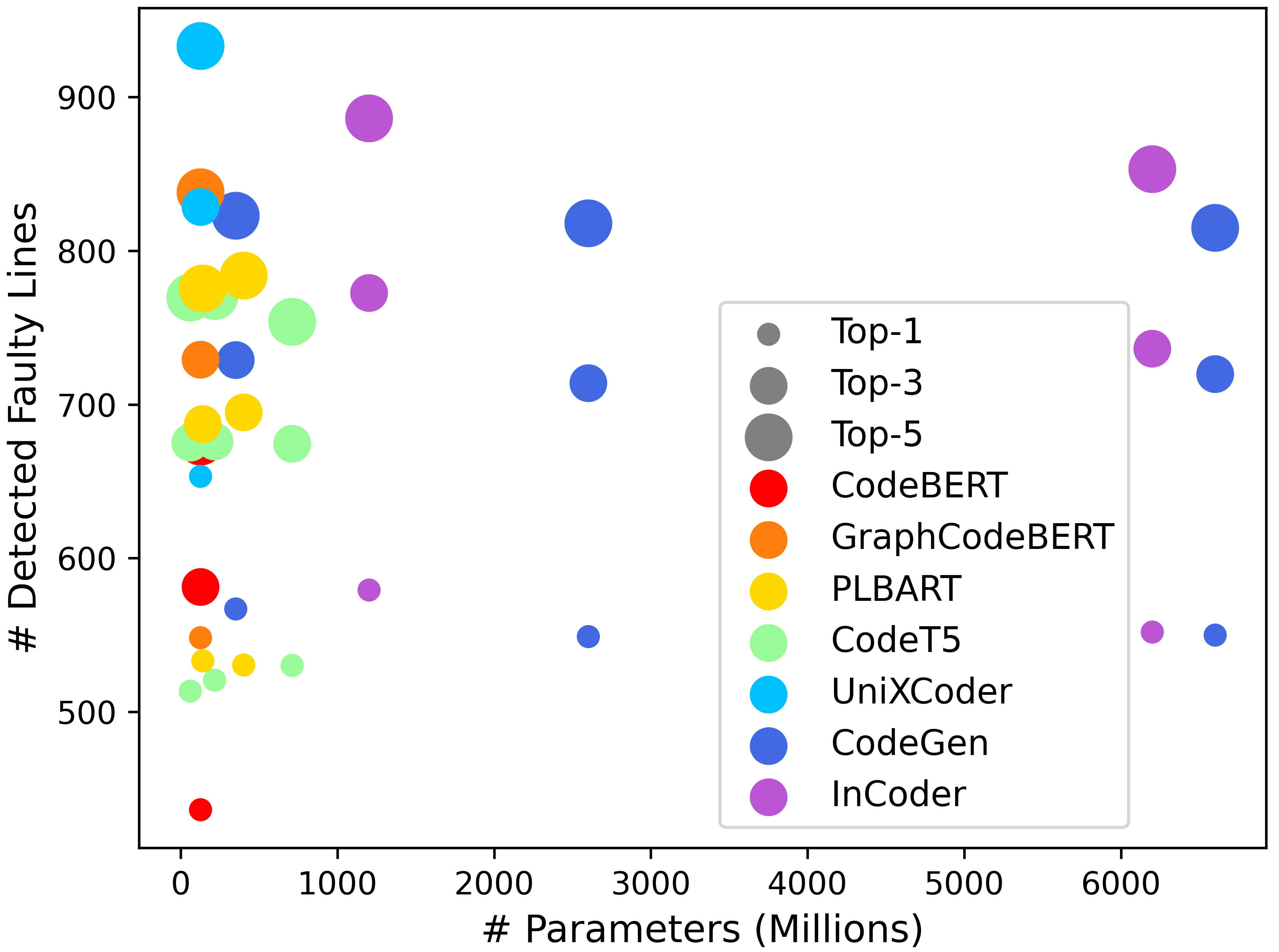}
    \caption{RQ1. FL performance based on model size}
    \label{fig:FL-model-size}
\end{figure}

\textbf{Model size.}
Figure~\ref{fig:FL-model-size} illustrates the performance of LLMCs with different model sizes. PLBART, CodeT5, CodeGen, and InCoder have been released with different parameter sizes by increasing the length of input/output sequences or hidden layers, as well as by stacking more layers on the same architecture. For PLBART, increasing the parameter size from 140M (base) to 400M (large) results in a slight performance improvement. For CodeT5, the performance of 60M (small) and 215M (base) is almost similar but decreases slightly when scaled up to 710M (large). CodeGen and InCoder also show a slight performance decrease as the model size increases. GraphCodeBERT and UniXcoder show impressive performance even with 125M parameters, a relatively small size. The FL method proposed in this paper does not generate new sequences from the decoder but rather outputs part of the input sequence as is, which may explain why the performance tends to decrease for more complex and larger models.

\begin{FindingBox}
\textbf{Finding 2:} In our FL approach, increasing the model size does not significantly improve performance. In particular, models with more than 500M parameters tend to perform worse. InCoder-1B model performs well though, due to its pre-training method, as highlighted in Finding 1.
\end{FindingBox}

To simplify the following experiments, we will now only use GraphCodeBERT-base (encoder only), CodeT5-small and UniXcoder-base (encoder-decoder), and CodeGen-350M and InCoder-1B (decoder only) models to compare FL performance.

\subsection{Comparison with SOTA FL Techniques}

\begingroup
\begin{table*}[t]
  \captionsetup{justification=centering, margin=3.1cm, width=0.9\textwidth}
  \caption{RQ2. Comparison of LLMCs fined-tuned by our approach with the state-of-the-art FL techniques}
  \label{tbl:RQ2-llmc-sota-comparison}
  \small
  \begin{tabular}{c|c|ccc|ccc}
    \toprule
    \multirow{2}{*}{Type} & \multirow{2}{*}{Model} & \multicolumn{3}{c|}{\textbf{Defects4J v1.2} $(N = 535)$} & \multicolumn{3}{c}{\textbf{Defects4J v2.0} $(N = 756)$} \\
    & & Top-1 & Top-3 & Top-5 & Top-1 & Top-3 & Top-5 \\
    \midrule
    Encoder only & \textbf{GraphCodeBERT}~\cite{guo2020graphcodebert} & 233.6 & 304.3 & 345.3 & 315 & 425 & 493.6 \\ \hline
    \multirow{2}{*}{Encoder-Decoder} & \textbf{CodeT5}~\cite{wang2021codet5} & 215.6 & 274 & 307.3 & 299.3 & 404.3 & 465 \\
    & \textbf{UniXcoder}~\cite{guo2022unixcoder} & \textbf{268} & \textbf{334.3} & \textbf{371} & \textbf{385.3} & \textbf{494.3} & \textbf{562.3} \\ \hline
    \multirow{2}{*}{Decoder only} & \textbf{CodeGen}~\cite{nijkamp2022codegen} & 232 & 297 & 338 & 335 & 432 & 485 \\ 
    & \textbf{InCoder}~\cite{fried2022incoder} & 245.3 & 319.6 & 361.6 & 334.6 & 455.3 & 525 \\ \hline
    \multirow{3}{*}{PEFT-based} & \textbf{LLMAO} (10-fold)~\cite{yang2024large} & 185 & 203 & 230 & - & - & - \\
    & \textbf{LLMAO} (new projects) & - & - & - & 133 & 145 & 166 \\ 
    & \textbf{LLMAO} (Recoder) & 126 & 157 & 176 & 154 & 187 & 201 \\ \hline
    \multirow{2}{*}{Spectrum-based} & \textbf{Ochiai}~\cite{abreu2006evaluation} & 72 & 133 & 167 & 59 & 147 & 196 \\ 
    & \textbf{Jaccard}~\cite{abreu2009practical} & 72 & 134 & 167 & 58 & 146 & 194 \\ \hline
    \multirow{2}{*}{Mutation-based} & \textbf{MUSE}~\cite{moon2014ask} & 102 & 146 & 188 & 117 & 220 & 289 \\ 
    & \textbf{Metallaxis}~\cite{papadakis2015metallaxis} & 92 & 176 & 221 & 139 & 273 & 341 \\ \hline
    \multirow{3}{*}{Learning-based} & \textbf{DeepFL}~\cite{li2019deepfl} & 135 & 230 & 299 & 144 & 312 & 404 \\ 
    & \textbf{XAI4FL}~\cite{widyasari2022xai4fl} & 165 & 279 & 331 & 165 & 330 & 417 \\ 
    & \textbf{TRANSFER-FL}~\cite{meng2022improving} & 208 & 306 & 361 & 275 & 433 & 500 \\ 
    \bottomrule
  \end{tabular}
\end{table*}
\endgroup

\textbf{RQ2. How effective are LLMCs compared to the state-of-the-art FL techniques?} Table~\ref{tbl:RQ2-llmc-sota-comparison} compares the FL performance of LLMCs fine-tuned with our approach and representative FL tools of each type when tested on Defects4J v1.2 and v2.0, respectively. Traditional (spectrum and mutation-based) studies and the learning-based XAI4FL compute the suspiciousness of lines from coverage data measured by executing the code with test cases (XAI4FL builds one model per defect). Meanwhile, in learning and PEFT-based studies, Defects4J data is divided for training and evaluation to train the models. DeepFL performs $N$-fold CV on defects within the same project in Defects4J, and TRANSFER-FL and LLMAO perform 10-fold CV across the entire data regardless of projects to address data leakage.

We re-train LLMAO, which takes source code as input, on the function-level data we curate. We perform 10-fold CV on Defects4J v1.2 for LLMAO (10-fold), which shows a similar level of performance reported in~\cite{yang2024large}. However, LLMAO (new projects), trained on Defects4J v1.2 and evaluated on v2.0, or LLMAO (Recoder), trained on Recoder data and evaluated on the Defects4J benchmark, show poorer performance compared to what the authors report. This could be considered for two reasons: first, since LLMAO is a model designed to estimate the location of defects in long programs at the project level, it may not be optimized for analyzing function-level data. Second, judging from the fact that the 10-fold case shows similar performance to that in~\cite{yang2024large}, data leakage may still occur even in 10-fold CV. Similarly, TRANSFER-FL may also have the possibility of data leakage since it also uses 10-fold CV (but there are cases where reproducing Defects4J defects is impossible and thus are considered to be unlocated faults: 29 in v1.2, 40 in v2.0).

The LLMCs fine-tuned with our approach show excellent performance even without using additional information about defects (such as coverage information or input-output examples). GraphCodeBERT, a pre-trained encoder-only type that uses structural features of code, shows similar performance to TRANSFER-FL. The performance of InCoder, a decoder-only model but pre-trained to use bidirectional context information of the input sequence, is slightly better than TRANSFER-FL. UniXcoder, an encoder-decoder type that utilizes structural information of code and was pre-trained in three types, performs 1.35, 1.12, and 1.08 times better than TRANSFER-FL (Top-1, 3, and 5, based on Defects4J v2.0).

\begin{FindingBox}
\textbf{Finding 3:} In the FL results on the Defects4J benchmark, UniXcoder, fine-tuned with completely separate training and evaluation data, demonstrates 1.35, 1.12, and 1.08 times (Top-1, 3, and 5) better performance compared to the SOTA technique that utilizes all relevant information (such as coverage information, input-output examples, etc.), despite not having any information about the defects.
\end{FindingBox}

\subsection{Performance on Various FL Benchmarks}
\label{sec:eval-various-benchmarks}

\begin{table*}[!t]
\caption{RQ3. FL performance of LLMCs fine-tuned by our approach on other benchmarks}
\label{tbl:RQ3-other-benchmarks}
\small
\centering
\begin{tabular}{c|ccc|ccc|ccc|ccc} 
\toprule
\multirow{2}{*}{LLMCs} & \multicolumn{3}{c|}{\textbf{HunamEval-Java}} & \multicolumn{3}{c|}{\textbf{QuixBugs} (Java)} & \multicolumn{3}{c|}{\textbf{CodeNet} (Python)} & \multicolumn{3}{c}{\textbf{CVEfixes} (C/C++)} \\ 
& Top-1 & Top-3 & Top-5 & Top-1 & Top-3 & Top-5 & Top-1 & Top-3 & Top-5 & Top-1 & Top-3 & Top-5 \\
\midrule
\textbf{GraphCodeBERT} & 77.3 & 105 & 123.6 & \textbf{12.3} & \textbf{16.6} & \textbf{20.6} & 7,864 & 9,042 & 9,698 & 33.8$^*$ & 38.2$^*$ & 43.2$^*$  \\ 
\textbf{CodeT5} & 55 & 83.3 & 95 & 9.6 & 13.3 & 17.6 & 7,343 & 8,549 & 9,086.6 & 35 & 41.6 & 47.2 \\ 
\textbf{UniXcoder} & \textbf{89.6} & \textbf{114} & \textbf{131.6} & 12 & 16.3 & 19.3 & \textbf{8,535} & \textbf{9,876.6} & \textbf{10,341.3} & \textbf{52.6} & \textbf{72.2} & \textbf{88.4} \\ 
\textbf{CodeGen} & 50 & 79 & 98 & 5 & 11 & 17 & 8,118 & 9,292 & 9,747 & 39.8 & 52.4 & 61.8 \\ 
\textbf{InCoder} & 56.3 & 89.6 & 106.3 & 10 & 13.6 & 18 & 8,213 & 9,375 & 9,898 & 40.8 & 60.8 & 70.6 \\  \cmidrule{1-13}
Total & \multicolumn{3}{c|}{164} & \multicolumn{3}{c|}{40} & \multicolumn{3}{c|}{12,000} & \multicolumn{3}{c}{238 (5-fold)} \\
\bottomrule
\multicolumn{13}{@{}l@{}}{* These are the results of CodeBERT since GraphCodeBERT does not support C/C++ DFGs and cannot be applied to C/C++ code as it is.} \\
\end{tabular}
\end{table*}

\textbf{RQ3. How well do LLMCs perform on various FL benchmarks?}
Table~\ref{tbl:RQ3-other-benchmarks} shows the performance on 4 benchmarks (HumanEval-Java, QuixBugs, CodeNet, CVEfixes) to investigate the generalizability of FL by LLMCs fine-tuned by our approach and their performance in a different domain, vulnerability detection. 

\textbf{Java benchmark.}
Since HumanEval-Java and QuixBugs are in the Java language, we evaluate them with the same LLMCs fine-tuned on the Recoder dataset as in the previous experiments. Jiang~\emph{et al.} discovered that Defects4J is included in CodeSearchNet~\cite{husain2019codesearchnet} and BigQuery data, which are used for pre-training LLMCs, and thus released HumanEval-Java to avoid data leakage~\cite{jiang2023impact}. For this data that the models have not seen during pre-training, GraphCodeBERT and UniXcoder, which use an encoder that utilizes bidirectional context information of the code, localize more faults, compared to decoder-only models. CodeT5, which also uses an encoder structure, localizes a similar degree of faulty locations as CodeGen, unlike in the previous experiments. QuixBugs is a benchmark that implements popular algorithms such as quick sort and bit count, and it is highly likely that it was used as pre-training data for LLMCs, regardless of the programming language. Nevertheless, GraphCodeBERT and UniXcoder, which can understand bidirectional context information, still show superior FL performance. Additionally, since the two benchmarks consist of short, function-level programs, CodeT5, which is a smaller model but has an encoder structure, performs similarly to decoder-only models.

\begin{FindingBox}
\textbf{Finding 4:} For short programs, whether used for pre-training or not, GraphCodeBERT and UniXcoder, which were pre-trained to utilize the structural information of code, perform well in localizing faults. Additionally, the smaller CodeT5 model shows similar performance to larger decoder-only models by utilizing its encoder structure.
\end{FindingBox}

\textbf{Python benchmark.} This benchmark is also primarily composed of short programs, and we divide 120,000 data points into 8:1:1 for fine-tuning, validation, and evaluation. For this benchmark, UniXcoder performs best, and unlike the Java benchmarks above, the decoder-only models perform slightly better than GraphCodeBERT. This is because among the pre-training data of LLMCs, the ratio of Python:Java is 9:1 for InCoder and 2.3:1 for CodeGen, whereas 1.54:1 for GraphCodeBERT and UniXcoder which use the same dataset. Although it has learned relatively less Python data, UniXcoder utilizes the bidirectional context information of the input code sequence, learned from the three types of pre-training methods, and correctly localizes 71.1\%, 82.3\%, and 86.1\% of faults in Top-1, 3, and 5, respectively.

\begin{FindingBox}
\textbf{Finding 5:} While the decoder-only models pre-trained on a large amount of Python data perform well, UniXcoder, which effectively pre-trained on a smaller amount of data, performs slightly better. This shows that while the amount of data impacts performance, an effective pre-training method has an even greater influence.
\end{FindingBox}

\textbf{C/C++ benchmark.} Since this benchmark has a small number of data points $(N = 1,190)$, we use 5-fold CV. GraphCodeBERT uses the code sequence and the corresponding DFG information as the input sequence, but since it does not support generating DFGs for the C/C++ language, the same type of CodeBERT is used instead. While UniXcoder and CodeBERT were pre-trained on the CodeSearchNet dataset, which does not contain C/C++ code, CodeT5, CodeGen, and InCoder were pre-trained on datasets that do contain C/C++ code. Nevertheless, UniXcoder, pre-trained to utilize the bidirectional information of the input sequence, performs best, localizing 22.1\%, 30.3\%, and 37.1\% of the vulnerabilities in Top-1, 3, and 5, respectively. Despite 1/3 of its pre-training data being in the C language, CodeT5 localizes a similarly small number of faults as CodeBERT.

\begin{FindingBox}
\textbf{Finding 6:} UniXcoder fine-tuned on the C/C++ vulnerability detection benchmark correctly localizes around 30\% of the vulnerabilities. This performance is about twice as good as CodeT5. Despite not including any C/C++ code in its pre-training data, UniXcoder learns to utilize information from the bidirectional context of code from other languages.
\end{FindingBox}

\subsection{Ablation Study}

\begin{table}[!t]
\caption{RQ4. FL performance of ablation of pre-training or line-numbering technique $(N = 1,291)$}
\label{tbl:RQ4-ablation}
\small
\begin{tabular}{c|c|ccc} 
\toprule
\multirow{2}{*}{\makecell{Fine-tuning \\ technique}} & \multirow{2}{*}{LLMCs} & \multicolumn{3}{c}{\textbf{Defects4J}} \\
& & Top-1 & Top-3 & Top-5 \\ 
\midrule
\multirow{3}{*}{\makecell{\textbf{w/o both} \\ pre-training and \\ line-numbering}} & CodeBERT-based & 11.3 & 19 & 30.6 \\
& CodeT5-based & 5 & 6.3 & 6.6 \\
& CodeGen-based & 12 & 13.3 & 13.3 \\ \hline
\multirow{3}{*}{\textbf{w/o pre-training}} & CodeBERT-based & 273.6 & 300.6 & 308.6 \\
& CodeT5-based & 281.6 & 304.3 & 320 \\
& CodeGen-based & 304.3 & 384.6 & 417.3 \\ \hline
\multirow{5}{*}{\textbf{w/o line-numbering}} & GraphCodeBERT & 318 & 419.3 & 479 \\
& CodeT5 & 311 & 396 & 438 \\
& UniXcoder & 395 & 524.3 & 600 \\
& CodeGen & 345 & 450 & 509 \\
& InCoder & 326 & 435 & 502 \\ \hline
\multirow{5}{*}{\makecell{\textbf{w/o sequence} \\ \textbf{generation} (SG)}} & GraphCodeBERT & 454.6 & 633.6 & 765.6 \\ 
& CodeT5 & 446 & 586 & 679 \\
& UniXcoder & 595.3 & 769.3 & 867 \\
& CodeGen & 531 & 715 & 828 \\
& InCoder & 532 & 741 & 851 \\ \hline
\multirow{5}{*}{\makecell{Fine-tuning with \\ \textbf{our approach}}} & GraphCodeBERT & 548.3 & 729.3 & 838.3 \\
& CodeT5 & 513.6 & 675.3 & 769.6 \\
& UniXcoder & \textbf{653.3} & \textbf{828.6} & \textbf{933.3} \\
& CodeGen & 567 & 729 & 823 \\
& InCoder & 579.3 & 772.6 & 886.3 \\ 
\bottomrule
\multicolumn{5}{@{}l@{}}{\makecell[l]{For `w/o both' and `w/o pre-training', we use the baseline models of each \\ type, which do not use any special pre-training techniques for code.}} \\
\end{tabular}
\end{table}

\textbf{RQ4. How does pre-training or line-numbering technique affect the FL performance of LLMCs?} 
Table~\ref{tbl:RQ4-ablation} shows the FL performance of the ablation study.

\textbf{(1) For w/o both}, we use LLMCs (from scratch) initialized with random weights (i.e. vanilla LLMCs without any pre-training for code) by directly inputting the original function sequences without line-numbering and training them to infer the sequence at fault locations. In this case, the trained LLMCs could only find less than 3\% of the faults. In particular, the CodeBERT-based model with an encoder outperforms the decoder-only model. We note that the size of the CodeT5-based model is 60M, which is about 3 times smaller than the CodeBERT-based model connected to a decoder (165M), and about 6 times smaller than the CodeGen-based model, resulting in lower performance.

\textbf{(2) For w/o pre-training}, we use LLMCs initialized with random weights, but train them with the input and output sequences using our line-numbering technique. Compared to (1), the FL performance improves by about 10 times (CodeBERT, Top-5) to 56 times (CodeT5, Top-1). From these results, we can see that our proposed line-numbering technique has a significant impact on FL.

\textbf{(3) For w/o line-numbering}, we fine-tune pre-trained LLMCs using the original function sequences without line-numbering and the fault location sequences, which is the most basic form of FL using LLMCs. Compared to (1), the FL performance improves by about 60 times (CodeT5) and 30 times (CodeGen), respectively. In addition, UniXcoder correctly identifies 46.5\% of the faults in Top-5 out of 1,291 methods with faults. This shows that the pre-training technique of LLMCs has a significant impact on FL.

\begin{FindingBox}
\textbf{Finding 7:} While pre-training of LLMCs is crucial for FL through sequence generation, our line-numbering technique with sequence generation is also important. In particular, even when trained from scratch, a certain level of FL performance is observed when using our approach.
\end{FindingBox}

\textbf{(4) For w/o sequence}, we use pre-trained LLMCs and fine-tune them using line-numbered input sequences of code, and predict only the line numbers for the faults as output, to examine the effectiveness of our approach. The experimental results show that the performance improves from 1.43 times (GraphCodeBERT, Top-1) to 1.7 times (InCoder, Top-3) compared to (3). UniXcoder identifies 67.2\% of fault locations in Top-5. This difference is due to the fact that in (3) a fault location is considered found if the faulty input line and the model's predicted sequence match after removing leading/trailing whitespaces from the predicted sequence, whereas in (4) line numbers are compared directly. However, even considering this difference, our line-numbering technique has a significant impact on FL.

\begin{FindingBox}
\textbf{Finding 8:} Even when generating only line numbers instead of whole line sequences, it still shows better performance compared to (3). As shown in Finding 7, line-numbering seems to be important and plays a special role as the tokens for the lines contain compressed information about their corresponding lines.
\end{FindingBox}

\textbf{(5) For our approach}, we fine-tune pre-trained LLMCs using the input and output sequences with line-numbering applied, incorporating all techniques. Compared to (4), the performance improvement for the decoder-only models is small, while the improvement for the models with an encoder structure is significant. The purpose of the decoder-only models is to predict the next token from the current input sequence. Therefore, since the starts of the output sequences in (4) and (5) are the same line numbers, they show similar performance. However, models with an encoder predict fault locations from bidirectional context information in the input sequence, resulting in a greater performance boost.

\begin{FindingBox}
\textbf{Finding 9:} When fine-tuning to infer both the line number and the sequence of the corresponding line together, LLMCs with an encoder structure that utilizes bidirectional context information from the code show an improvement in FL performance ranging from 1.1 times (UniXcoder) to 1.2 times (GraphCodeBERT) in Top-1, compared with those without SG.
\end{FindingBox}

\section{Threats to Validity and Limitations}

\paragraph{Internal} It is possible that the pre-training data of LLMCs included the benchmarks we use. This implies that the models might have been evaluated on data that they had already seen during pre-training~\cite{jiang2023impact,xia2022less}. As mentioned in Section~\ref{sec:eval-various-benchmarks}, LLMCs could have been pre-trained on Defects4J data, which are commonly used benchmarks in FL, through CodeSearchNet and BigQuery. Moreover, given that CodeGen and InCoder were pre-trained on 300 GB and 52 GB of Python data, respectively, there is a significant likelihood that our Python benchmark data was included in their training datasets. However, it is important to note that LLMCs tend to forget the knowledge acquired during pre-training in the fine-tuning phase~\cite{baudry2021software, shao2022overcoming}. Therefore, we believe that the impact of data leakage on our evaluation is minimal. Additionally, we evaluate LLMCs using the HumanEval-Java benchmark. Given the timing of this benchmark's translation from Python into Java, it is highly unlikely that the pre-training datasets of the seven LLMCs used in this study included this data, ensuring a fair comparison.

Furthermore, we do not consider any filtering or re-ranking strategies for the FL candidates generated by LLMCs. This indicates that while the fine-tuned LLMCs identify a considerable number of faults compared to the SOTA techniques, there is still potential for improved performance through these methods. We are currently investigating a re-ranking strategy that utilizes the probability of patches generated by LLMCs, and expect significant improvements with contributions from subsequent researchers.

\paragraph{External} Due to the constraints of available resources, we employed LLMCs with a size of less than 6B that can be fine-tuned on four NVIDIA GeForce RTX 4090 GPUs. Consequently, extremely large language models such as Codex-12B~\cite{chen2021evaluating}, AutoCoder-33B~\cite{lei2024autocoder}, CodeLlama-70B~\cite{roziere2023code}, GPT3-175B~\cite{brown2020language}, and GPT4~\cite{achiam2023gpt}, which contains 1.7 trillion parameters and is challenging to run under typical circumstances, were excluded from our experiments. These models, having been pre-trained on more extensive datasets, inevitably pose a data leak threat. Although model size does not significantly impact performance, as indicated by our findings, the results for extremely large language models may differ, warranting a comparison. Future work involves enhancing our study by building a system with more substantial resources and conducting experiments with larger models. As this paper presents pioneering research on fine-tuning LLMCs for the FL domain, we expect further improvements by subsequent researchers.

\paragraph{Construct}
We fine-tune LLMCs to locate defects in relatively short, function-level programs due to the structural limitations of LLMCs. In contrast, previous studies often analyze long, project-level programs and calculate fault suspiciousness scores for numerous individual lines. To ensure a fair comparison, we extract only those lines within the faulty functions (as used as inputs to LLMCs) from the line-level suspiciousness scores calculated by previous studies. In other words, while previous studies utilize project-level information, LLMCs leverage only information specific to the faulty function. To address this discrepancy, Yang~\emph{et al.}~\cite{yang2024large} employ a sampling strategy to estimate fault locations in long, project-level programs. Incorporating this strategy into our approach could potentially enhance performance by utilizing project-level information. This integration, along with further research leveraging sufficient resources, remains part of our future work.

\section{Related Work}

Since LLMCs are already discussed in detail in Section~\ref{sec:preliminaries}, this section focuses mainly on traditional and learning-based FL techniques. Traditional FL techniques predict fault locations by utilizing program spectrum information such as execution paths, data dependencies, and program outputs obtained through testing~\cite{harrold2000empirical}. These techniques are based on the premise that defects are more likely to exist in those lines of code executed by failing test cases. 

SBFL calculates fault suspiciousness scores for each code element using various metrics based on the execution flow and pass/fail results of test cases. In SBFL, the number of test cases that execute a particular line or function and their outcomes (pass or fail) are counted. Research in this area focuses on metrics that calculate fault suspiciousness scores with high accuracy, such as Tarantula~\cite{jones2002visualization}, Ochiai~\cite{abreu2006evaluation}, Dstar~\cite{wong2013dstar}, Barinel~\cite{abreu2009spectrum}, GP13~\cite{yoo2012evolving}, Nashi1~\cite{naish2011model}, and Jaccard~\cite{abreu2009practical}, to name a few.

MBFL is a popular traditional FL technique. Although it requires additional cost for mutation compared to SBFL, MBFL is expected to deliver better FL performance. Traditional mutation techniques (MUJAVA~\cite{ma2006mujava}, Major~\cite{just2014major}, and PITest~\cite{coles2016pit}) allow mutation at the Java bytecode level or bit level along with the Java compiler, and mutation operators have to be defined manually. DeepMutation~\cite{tufano2020deepmutation} attempts to generate mutations closer to actual errors using a recurrent neural network. Meanwhile, LEAM~\cite{tian2022learning} attempts to generate syntax-driven mutations through grammar rules and a Transformer-based model.

LBFL typically trains deep learning models using data from traditional FL techniques such as SBFL and MBFL, while also incorporating additional information as input to the models. For instance, XAI4FL~\cite{widyasari2022xai4fl} applies explainable AI to existing spectrum-based features to determine which lines are associated with test results, using this information to calculate line-level suspiciousness scores. DeepFL~\cite{li2019deepfl} predicts function-level suspiciousness scores by incorporating complexity-based fault proneness and textual similarity information in addition to spectrum-based and mutation-based features obtained from functions. Lou~\emph{et al.}~\cite{lou2021boosting} observe that converting coverage analysis results into numerical or boolean values loses significant information. Therefore, they propose GRACE which generates graphs from the testing information and learns these graphs using a GCN (Graph-based Convolutional Network) model. DeepRL4FL~\cite{li2021fault} predicts function-level suspiciousness scores by learning code coverage, dependency representation, and source code representation using a CNN model. TRANSFER-FL~\cite{meng2022improving} improves FL by predicting bug types for each line using separate models, employing these as semantic-based features. It integrates spectrum-based, mutation-based, and semantic-based features into a model to predict line-level suspiciousness scores. FixLocator~\cite{li2022fault} implements models to predict both function-level and line-level faults, identifying all functions and lines that need modification for a single bug. It constructs function-level data (such as function information, execution information, and co-change relations) and statement-level data (such as abstract syntax trees and variable information), learning these using a GCN model. GMBFL~\cite{wu2023gmbfl} creates graph-based representations of static and dynamic analysis information of programs, learning these representations using a GGANN (Gated Graph Attention Neural Network) model.

\section{Conclusion}
To the best of our knowledge, this paper proposes the first sequence generation approach to fine-tuning LLMCs for FL and studies its impact on three representative types of LLMCs. We fine-tune 13 LLMC instances and evaluate their FL performance on four benchmarks (three for Java and one for Python), as well as one benchmark for evaluating their vulnerability detection performance in C/C++. Experimental results show that UniXcoder, pre-trained with three different types, demonstrates the best performance. LLMCs utilizing the encoder architecture to exploit structural properties of programs or pre-training methods that consider bidirectional context exhibit superior performance in FL. Specifically, when evaluated on the Defects4J benchmark, UniXcoder improves its Top-1 metric by approximately 1.7 times compared to not using our approach and outperforms the SOTA technique, TRANSFER-FL, by 1.35 times. In conclusion, our experimental results suggest that LLMCs have significant potential in the field of FL and warrant further research.
\begin{acks}
\end{acks}

\bibliographystyle{ACM-Reference-Format}
\bibliography{references}
\end{document}